# Biophysics in Africa: challenges, priorities, and hopes

Tjaart P. J. Krüger

*Department of Physics, University of Pretoria, South Africa*

Email: tjaart.kruger@up.ac.za

*This report is a serious call to scientists, innovators, investors, and policymakers to invest in the development of biophysics in Africa. The complex problems of our day demand multidisciplinary approaches, and biophysics offers training in much-needed multi- and cross-disciplinary thinking. Biophysics is a research field at the forefront of modern science because it provides a powerful scientific platform that addresses many of the critical challenges humanity faces today and in the future. It is a vital source of innovation for any country interested in developing a high-tech economy. However, there is woefully little biophysics educational and research activity in Africa, representing a critical gap that must be addressed with urgency. This report suggests key research areas that African biophysicists should focus on, identifies major challenges to growing biophysics in Africa, and underscores the high-priority needs that must be addressed.*

## Introduction and Motivation

Since the COVID-19 pandemic, many governments have expressed the need for Africa to be able to make its own therapeutics and vaccines. The first step for that to happen is investing in the basic and applied sciences and engineering research, and that especially means biophysics.

Why biophysics? This unique interdisciplinary field brings our understanding of biological processes to an unprecedently detailed level. Only when we understand nature's processes at a sufficiently deep level can we make reliable predictions and obtain sustainable technological solutions.

This is realised by numerous pharmaceutical companies where biophysics forms an indispensable component of drug discovery [1]. Dr. Martin Friede from the World Health Organization's Initiative for Vaccine Research took it a step further by stating, "It is impossible to develop the next generation of vaccines without biophysics" [2].

Consider Structural Biology, a subdomain of biophysics that aims to resolve and study the structure and dynamics of biological macromolecules such as proteins – the molecular machines of biological cells. Knowing the protein structure at the atomic level has enormous commercial potential in areas such as industrial enzymology and drug discovery. A fully resolved protein structure enables us to engineer proteins that can make new chemicals and to design molecules that interfere with the life-giving reactions of harmful pathogens or pests (i.e., drugs and pesticides). Structural Biology is therefore an important step to the global economic success of a country. It is particularly telling that over 80 Nobel prizes have thus far been awarded to the field of Structural Biology: 70 prizes for scientific discoveries and 11 prizes for experimental methods that enabled these discoveries [3].

Biophysics is not only concerned with scientific research. An integral component of scientific discovery in biophysics involves technological development. Innovative experimental and computational methods pave the way for new scientific discoveries and provide practical solutions across the broad domain of biological sciences. In this sense, biophysics is more than a basic science that feeds innovation, because innovation is an integral part of research in biophysics.

Biophysics revolutionised medical research and technology in the 20[th] century. It provided both the tools and the understanding for treating various diseases. These developments are accelerating in the 21[st] century. Biophysics addresses not only human health challenges but also plant and animal health. By understanding the minutiae of photosynthesis through decades of scientific research, rice and soy plants were recently engineered with 20–30% enhanced crop yield [4–6].

Biophysics research features in various aspects of the global effort to combat climate change. An important area of research and technological development in this regard is biomimetics, which originates from biophysics [7]. The design of environmentally friendly materials such as biodegradable plastics is one example. Another example is how biomimetics offers a useful perspective in addressing food security and sustainable energy, two of the great challenges of our time: we can gain inspiration from the efficacy and adaptability of photosynthetic organisms to produce food or fuel from sunlight using materials that are very abundant in nature (i.e., inexpensive and scalable) [8]. In addition, meeting food, water, and energy demands is not limited to mankind, but it is a basic need of essentially every cell of every living organism. It is therefore prudent to investigate how other living organisms meet these demands at various levels.

Quantum Biology is a new, emerging research field with enormous potential for science and technology. This field of research investigates how biological organisms use the principles of quantum mechanics to gain a physiological advantage in executing their physiological functions [9, 10]. During the past few years, several research programmes focussing on Quantum Biology have been launched across the world [11]. It is important that Africa actively contributes to the development of this promising field of research. Applications of Quantum Biology could impact many technologies, such as energy, environment, health, sensing, and information technologies [9, 10, 12]. Learning from life will not only lead to new technologies but also to new fundamental insights in physics, chemistry, and biology. For example, in the medical field, it is known that light enhances wound healing and effectively treats different types of cancer, and when applied to the brain it can have a range of physiological effects such as improved attention, memory, executive function, and rule-based learning [12]. Identifying how quantum effects might play out in the brain could offer a completely new way of imagining medical intervention beyond the purely chemical.

The term "century of biology" was coined for the 21$^{st}$ century in the context of biotechnological development [13] to address several critical global challenges. Biophysics plays an indispensable role both in establishing the crucial scientific basis and in bridging the gap between science and technology.

A strong and diverse biophysics research and commercial sector is essential for the success of the African economy. The importance of the bioeconomy has been recognised by numerous countries. For example, the UK [14], EU [15], USA [16] as well as South Africa [17] have formulated strategies to move away from the traditional industrial base and instead develop a strong bioeconomy. Notably, biophysics is an indispensable component of these bioeconomy strategies.

## Key research areas requiring biophysicists

The most important Sustainable Development Goals (SDGs) of the United Nations that biophysics addresses directly are 2. Zero Hunger, and 3. Good Health and Wellbeing. Biophysics research contributes to several additional SDGs, albeit usually indirectly, viz. 1. No Poverty; 8. Decent Work and Economic Growth; 9. Industry, Innovation and Infrastructure; 12. Responsible Consumption and Production, 13. Climate Action; 14. Life below Water, and 15. Life on Land. In addition, the development of biophysics on the African continent requires a strong commitment to quality education (SDG 4).

Zero Hunger (SDG 2) is addressed through biophysics research and development in agribusiness and food security, which involves important aspects such as food nutrition and the understanding, prevention, and treatment of animal and plant diseases. An improvement in food security is directly linked to Good Health and Wellbeing (SDG 3), which also involves the large research field of human health and medicine. These two SDGs are a particularly pressing need for Africa, but biophysics offers an abundance of opportunities to address them in tangible ways.

1. Medicine

We wish to focus specifically on diseases that constitute the most significant health, social, and economic burden to the African continent. These include (i) poverty-related diseases such as HIV/AIDS and tuberculosis, which kill millions of people annually, (ii) neglected tropical diseases that affect in the order of 400 million people on the continent according to the World Health Organization, (iii) malaria with an annual mortality rate of about half a million African people, and (iv) cancer, for which the mortality rate increases every year and is predicted to reach ca. 1.4 million annual deaths in Africa by 2040 [18].

Most of these mortality cases can be linked to the patient having limited access to treatment or the inability to afford the treatment. African countries therefore have a desperate need for robust, cost-effective diagnostics and low-cost innovations to address local needs – and biophysics plays a crucial role in the development of these technologies.

Another key area of research is the development of drugs and vaccines for which research in Structural Biology is indispensable. By resolving molecular structures of macromolecules, Structural Biology provides the tools to understand the molecular basis of diseases, which guides the rational design of new drugs and the optimisation of existing medicines.

Besides Structural Biology, other key areas of biophysics research in the health sector include biosensing and quantum biology to enable sensitive diagnostics, biophotonics for numerous applications including light therapy and sensitive diagnostics, cost-effective imaging solutions, and computational approaches to complement experimental work and deepen our understanding of diseases.

## 2. Agribusiness/food security

Biophysics can also contribute significantly to agribusiness in several ways, in particular by scientific and technological solutions to improve plant health. Growing food insecurity and sustained malnutrition are a major concern in the developing world. The rapidly growing food demand is due to the combination of a growing African population and a reduction in fertile farmland. This requires drastic agricultural intensification, which means that plant health becomes an increasingly important demand every year.

Currently, at least half of agricultural loss occurs due to biotic or abiotic stressors. Biotic stressors are stress factors of a biological origin, for example, pathogens, insects, fungi, parasites, worms, and weeds. Abiotic stressors are non-biological factors such as non-optimal soil salinity, nutrient deficiency, drought, extreme temperature, and excess light.

Early plant disease detection is an emerging area of research, constituting non-invasive methods – typically remote sensing technologies – that enable early, pre-symptomatic diagnosis of plant stress [19, 20]. These methods enable the farmer to treat diseases or optimise abiotic factors at the earliest stages, which can be several days before the plants would show symptoms that are observable by the eye. Early treatment curbs the spread of diseases, increases the chances of successful treatment, and reduces the resources required for treatment. The non-invasiveness of these methods also enables precision agriculture and plant phenotyping for resistance breeding [21, 22]. Remote sensing includes numerous promising spectroscopy-based methods such as hyper- and multispectral imaging and pulse-amplitude-modulation fluorometry. Owing to their deep understanding of spectroscopy, modelling, and device development, biophysicists are apt to enhance the sensitivity of these technologies, devise ways to relate spectroscopic changes to particular stress factors, and translate the detected signals between different environments (e.g., from indoor to outdoor) and across different scales (e.g., from the leaf to the canopy level). This is a largely unexplored area of research but crucial for maintaining crop productivity and food security.

Another promising area of biophysics research is to provide a basis for finding alternative treatments for plant diseases. Reducing chemical use for pest management is an urgent need in Africa for cost, food safety, and environmental sustainability. Key problems of the use of pesticides and fungicides are the growing resistance of pests and fungi, and their toxicity to humans, animals, and the environment. We therefore urgently need to develop alternative ways to enable more accurate use of fungicides in the short term and explore less toxic alternatives in the long term. An example is to control spore dispersal from fungi, which can only be done when understanding the mechanics of fungal dispersal [23]. Again, biophysicists are needed to provide such a mechanistic understanding. This is one of numerous underexplored areas of research.

Biophysics is also paramount to obtaining a deep understanding of the complex photosynthetic process. The onset of biotic and abiotic stressors triggers a series of photoprotective mechanisms. It has been demonstrated that the genetic modification of some of these mechanisms can significantly improve crop yields [4–6].

The biophysics research methods of relevance to agriculture are similar to those needed in medicine, namely, biosensing, quantum biology, biophotonics, imaging, and computational approaches. Structural Biology is equally important. In addition to explaining basic life processes, structural techniques are routinely employed in the pharmaceutical industry, agrochemical industry, and biotechnology communities around the world in support of efforts to understand molecular disease mechanisms, the rational design of pesticides, herbicides, small molecule and biologic medicines, and in optimising and designing biocatalysts.

## Major challenges to growing biophysics in Africa

The best way to grow and establish biophysics on the continent is to create adequate opportunities for state-of-the-art research on home soil. The major challenges to this goal are discussed here. It is important to note that these challenges feed one another. In other words, addressing one requires addressing them all.

1. **Vastly inadequate infrastructure and resources**

All research and development require appropriate infrastructure and resources. This is even more so for biophysics research operating at the forefront of science and technology. There are a handful of research centres scattered across Africa that house relevant infrastructure [24]. This is a good start but undoubtedly markedly insufficient. Most African countries do not have even basic equipment for biophysics research, while the equipment hosted by the rest of the countries is vastly inadequate [24]. The severe lack of equipment is a very demotivating factor for aspiring biophysicists on the continent.

Acquisition of equipment is only one side of the coin. Equally important is the need to maintain technical infrastructure by equipping our own people and providing sufficient funds. It has happened too often that state-of-the-art specialised equipment gets wasted because of inadequate resources to sustain it – due to a lack of expertise or funds for maintenance or both.

Consider as an example the infrastructure required for Structural Biology. Determining the structure of biological macromolecules requires the establishment of a workflow that includes the ability to prepare the material, test its functionality, obtain the data necessary for structure determination, process this data, and interpret the outcome. Both X-ray crystallography and cryo-electron microscopy lead to directly interpretable, near-atomic-resolution visualisations of biomolecular structures and are currently the most widely used structure determination techniques. The value of structural insights is recognised internationally to the extent that industries as well as governments abroad have invested billions in building and staffing shared, large-scale, centralised infrastructure for Structural Biology. In comparison, due to the high cost of the technology and the critically scarce skills required to operate such equipment, only limited structural investigations are possible at select sites in Africa, all of which are currently in South Africa. The technology and thus critical insights remain elusive to both local industry and academic researchers. Where resources have been committed, appropriate equipment and skills have been spread over many sites and this has meant that a productive critical mass that could lead to development and innovation has never been established. Trained students have in general not been retained and many have found employment in the field abroad, where they have been highly successful.

It is also important that one or more of the societal activities in which structural biology is needed must exist in a country interested in developing this field of research. For example, there should be companies researching novel agrochemicals, medicines, or industrial enzymes for which protein structural information is a *sine qua non*. Given the poor state of development of the discipline in Africa, it is unlikely that entrepreneurs will invest without substantial government intervention.

## 2. Very low critical mass

The present state of affairs is that very few students and research scientists in Africa venture into biophysics. One major reason is a lack of awareness of the importance of this field of research. This leads to limited funding opportunities supporting biophysics research and development, which, in turn, discourages scientific work in this area.

Another major reason for Africa's low critical mass in biophysics is the exodus of skilled scientists. Most Africans interested in biophysics study abroad and do not return to Africa, while most of those who returned to their home countries have remained in biophysics for short periods. The primary reason for this is the severe shortage of infrastructure and resources for biophysics research. These scientists have the necessary knowledge and skills but they lack the capacity to execute the research. Opportunities are urgently needed to support and help these scientists to excel in their research.

## 3. Limited educational, training, and mentorship opportunities in Africa

Going hand-in-hand with the previous two challenges is the need to educate, train, and mentor our current and aspiring biophysicists in Africa. Only a few African universities offer biophysics courses and even fewer offer biophysics degrees. In addition, general and specialised biophysics schools and workshops in Africa are organised too infrequently. Combined with education and training is the need for mentorship to encourage and nurture aspiring and established biophysicists on the continent.

# High-priority future needs

## 1. Capacity building

An earnest investment in educational opportunities is a low-hanging fruit for the growth of critical mass and knowledge in biophysics. This must be done through the development of biophysics curricula and the hosting of general and specialised biophysics schools, workshops, seminars, and expert lectures. Biophysics programmes and degrees would need to be established as a pipeline in developing curricula along both academic and vocational lines. Both Africans and non-Africans can help significantly to address these needs. In this regard, the International Union for Pure and Applied Biophysics (IUPAB) and the Biophysical Society (BPS) have ample resources that can be tapped into.

The development of biophysics research should be a natural outflow of biophysics education and training. Again, support from IUPAB and BPS as well as numerous other international societies would be of immense help, for example, to bring international experts to Africa through the organisation of workshops and conferences. Collaboration with well-established biophysicists in other continents through multinational research programmes and consortia is an excellent way to boost research quality and opportunities. This becomes a realistic opportunity when African researchers strive for excellence.

Lastly, the severe lack of awareness of biophysics on the continent must additionally be addressed through public awareness activities such as popular science literature, news reports, science festivals, roadshows, and school visits and demonstrations. In general, the profile of scientists must be raised in the public eye. They are the people expending great effort in training the next generation of leaders and developing innovative technological solutions. If scientists – and biophysicists in particular – could be elevated to the same level as sports stars, this would immediately attract significant attention from the public and governments. In addition, if scientists do not actively define their role in society, their relevance will be determined by society – and this will be a vastly underappreciated role.

## 2. Investment in infrastructure and equipment

As motivated above, the acquisition and maintenance of modern infrastructure and equipment is key to the development of biophysics research and innovation. Funding for this requires governmental support, which should grow through policy development and high-level discussions with governments convincing them of the need to support the work of African biophysicists, build the necessary infrastructure, and encourage African industries to invest in the bioeconomy strategy.

Governing bodies and investors must make adequate funding available for the procurement of necessary facilities for biophysics research. Funding incentives should also be provided to researchers to establish and develop biophysics research in important areas. To this end, governments may develop multiple-department initiatives to support the work of biophysicists. They should incentivise our universities to build infrastructure in all the fields that support biophysics and make funding available for basic and advanced equipment.

African home countries need to invest in their own research. Currently, the weakest link is the fact that we get most funding from outside Africa and no or very limited buy-in from our own continent. Africans must be convinced that their support is indispensable.

Investment in infrastructure and human capacity development must be seen for what it is: an investment – not for a limited number of elite persons but for the country and ultimately for the whole continent! A growing body of expertise will attract industrial development, which, in time, will inevitably lead to direct foreign investment and the development of intellectual property and products. Consider as an example the study of protein structure. Proper investment in the development of infrastructure and scientists to do cutting-edge Structural Biology research will enable the development of local industries concerned with drug discovery and development, advanced agrochemicals, and fourth-generation industrial biotechnology.

Biophysics research depends on a very broad spectrum of experimental techniques, and it is therefore impossible to house all the necessary equipment on the African continent. But it is also unnecessary to try and collect all types of equipment. Firstly, we must be selective in our focus, specifically addressing the key research areas stated above. Secondly, we must follow the example of European countries that similarly do not house all the necessary equipment but, instead, form consortia to share expensive equipment, which can also be accessed by scientists from non-member countries.

### 3. Low-cost innovations to address local needs

Although the importance of acquiring and maintaining expensive equipment for state-of-the-art biophysics research and development cannot be understated, a particularly pressing need for Africa is to find inexpensive technologies for the vast majority of its people who cannot afford expensive solutions. In this regard, it is important to note that for most applications, only a dedicated technology is needed, not a versatile one. This requirement may significantly decrease the price of the technology. Connected with this is the need to develop methods that are specific to particular contexts. Such affordable solutions require innovative thinking.

Consider as an example a quantum light imaging device to improve the resolution of medical images for people living in remote areas. This technology is out of place for its target group because, firstly, such equipment is very expensive; secondly, it requires a well-isolated (vibration-free) environment and reliable electricity supply; and, thirdly, it requires highly skilled staff to operate and maintain. Instead, a significantly cheaper instrument can be used to acquire an image at a lower resolution, after which machine learning techniques can be employed to optimise the image resolution.

Another example of an inexpensive innovative instrument is a homebuilt hyper- or multispectral camera, which can be a few orders of magnitude cheaper than state-of-the-art commercial ones. Such a camera can be built using a 3D printer and Raspberry Pi kit, the latter of which is then used to control inexpensive camera sensors and filters. Running the output through a machine-learning algorithm can again improve the image and spectral resolution. The cost of this instrument can be cut further when dedicated to a specific application. Possible applications are diverse and may include the sensing of particular stressors in plants, drug sorting, detection of tainted drugs, diagnosis of traditional medicines, food diagnosis to determine its safety for consumption (e.g., detection of pesticides, rot, or diseases), or investigation or detection of plastics.

These examples highlight the importance of translating scientific work from the laboratory to society by finding inexpensive, dedicated solutions. This is in line with the World Health Organization's set of criteria for ideal diagnostic test development based on the acronym REASSURED, which refers to **R**eal-time connectivity, **E**ase of specimen collection, **A**ffordable, **S**ensitive, **S**pecific, **U**ser-friendly, **R**apid and robust, **E**quipment-free or simple, and **D**eliverable to end-users.

## Synergies with neighbouring fields

The broad scope of biophysics demands a broad range of experimental and modelling approaches. Even within a focussed area of biophysics, numerous experimental and modelling approaches are often used to obtain a deep, integrative understanding of the complex system at hand. Therefore, biophysics has synergy with many other fields of physics.

Adopting a broad definition of biophysics here, biophysics has a strong overlap with many other disciplines such as biochemistry, biocomputing, biomathematics, biomedical engineering, biotechnology, botany, chemistry, crystallography, genetics, genomics, molecular biology, neuroscience, oceanography, pharmacology, physiology, structural biology, synthetic biology, systems biology. Professional African Societies for many of these disciplines already exist and biophysics initiatives must cooperate with these societies [24].

Cross-pollination of biophysics with the various subdisciplines of physics and the other related scientific disciplines is strongly recommended because this encourages lateral, cross-disciplinary thinking.

## Conclusion and Perspectives

Africa, with its wealth of human capital, has enormous potential to revolutionise the continent for the welfare of its people and the rest of the world. A strong investment in biophysics must form a vital component to reimagine its economic growth and increase its prosperity. The reason is that the complex world that we live in demands multidisciplinary approaches – and biophysics is by definition a multidisciplinary science that is strongly rooted in the useful value system of physics, endowing its adherents with critical thinking and problem-solving skills. The core questions of our day may come from one discipline, but the solutions are often found from its integration with other disciplines, and the integration of physics to resolve questions in biology is a beautiful example of this.

African universities should now begin to craft cross-disciplinary degrees – at the undergraduate and postgraduate levels – because it is cross-disciplinary research that is going to transform science in the decades to come. Biophysics is again an excellent example of such a cross-disciplinary approach.

To put Africa on the global biophysics maps, it is essential to establish multinational research programmes, consortia, and training events across the continent. There are already a number of exemplary initiatives. They must be sustained and inspire the development of many more initiatives.

## Acknowledgements

We are grateful for contributions from the following people:

- Trevor Sewell (Department of Integrative Biomedical Sciences, University of Cape Town, South Africa)
- Nana Engo Serge Guy (Department of Physics, University of Yaoundé I, Cameroon)
- Kayode A. Dada and Fatai A. Balogun (Centre for Energy Research and Development, Obafemi Awolowo University, Ile-Ife, Nigeria)
- Kelvin Mpofu (Council for Scientific and Industrial Research, South Africa)
- Betony Adams (Department of Physics, Stellenbosch University, South Africa, and The Guy Foundation)
- Emmanuel Nji and Daouda A.K. Traore (BioStruct-Africa)
- Raymond Sparrow and Thomas Franke (School of Engineering University of Glasgow, UK)